 \documentclass[twocolumn]{aastex631}

\shorttitle{Light curve analysis of V1723 Sco}
\shortauthors{I. Hachisu \& M. Kato}

\begin{document}

\title{A multiwavelength light curve analysis of the very fast nova V1723 Sco}


\author[0000-0002-0884-7404]{Izumi Hachisu}
\affil{Department of Earth Science and Astronomy,
College of Arts and Sciences, The University of Tokyo,
3-8-1 Komaba, Meguro-ku, Tokyo 153-8902, Japan}
\email{izumi.hachisu@outlook.jp}

\author[0000-0002-8522-8033]{Mariko Kato}
\affil{Department of Astronomy, Keio University,
Hiyoshi, Kouhoku-ku, Yokohama 223-8521, Japan}




.


\begin{abstract}
We have analyzed multiwavelength light curves of the very fast
nova V1723 Sco, based on our fully self-consistent nova explosion models.
The time-stretching method gives the distance modulus in the $V$ band of
$(m-M)_V = 15.3\pm 0.2$.  Then the absolute $V$ magnitude reaches 
$M_{V,\rm max}= m_{V, \rm max} - (m-M)_V= 6.77 - 15.3\pm 0.2 = -8.5\pm 0.2$.
Using our fully self-consistent nova outburst model combined 
with the optically thick winds on a $1.25 ~M_\sun$ white dwarf accreted by
a mass accretion rate of $\dot{M}_{\rm acc}=1\times 10^{-9}
~M_\sun$ yr$^{-1}$, we successfully reproduce the overall $V$
light curve with a free-free emission model as well as the supersoft X-ray
light curve with a blackbody approximation model.
The epoch of the first GeV gamma-ray
detection is almost coincident with the epoch of our model $V$ peak.
This supports the shock formation mechanism that a strong shock arises
soon after the optical $V$ maximum far outside the WD photosphere.
We conclude that the shocked shell is optically thin.
\end{abstract}





\section{Introduction}
\label{introduction}



A classical nova is triggered by an unstable hydrogen burning 
on a mass-accreting white dwarf (WD) in a binary
when a hydrogen-rich envelope accumulates mass up to a critical value
\citep[e.g.,][]{spa78, sio79, nar80, ibe82, pri95k}.  
The WD envelope expands to a giant star size and massive winds
emerge from the WD photosphere.  

Recently, realistic nova optical $V$ light curve models have been calculated 
in which optically thick winds are included in a fully self-consistent way 
\citep[e.g.,][]{kat22sha, kat25hs, kat26hs}. These are useful tools 
to determine the WD mass and mass accretion rate to the WD 
by fitting their optical $V$ and supersoft X-ray light curves 
with observation.  The WD mass, mass accretion rate, peak $V$ brightness,
and $t_2$ (or $t_3$) time are fundamental values to understand the nova
physics and binary evolution, where $t_2$ (or $t_3$) is the 2 (or 3) mag decay
time from the $V$ peak.
In the present paper, we determine these values of the very fast nova
V1723 Sco, based on our fully self-consistent nova explosion models.


\begin{figure*}
\epsscale{0.9}
\plotone{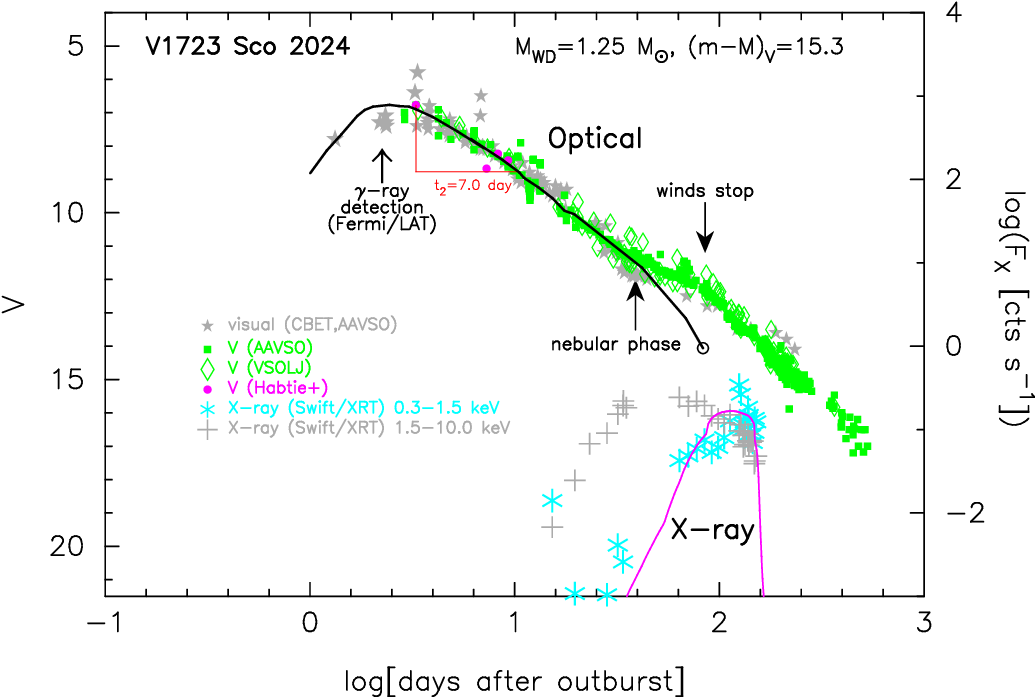}
\caption{
Summary of the optical $V$ (including visual and unfiltered $V$), and X-ray
(0.3-1.5 keV and 1.5-10.0 keV) light curves of V1723 Sco for both models
and observations.  The visual and $V$ data are taken from AAVSO,
VSOLJ, and \citet{habtie24}.  The visual and
unfiltered $V$ data are also taken from CBET No.5346 \citep{pearce24}.
The X-ray count rates are from the Swift website \citep{eva09}.
We add theoretical $V$ (black line) and X-ray (0.3-1.5 keV; magenta line)
light curves based on \citet{kat26hs}'s fully self-consistent nova outburst
model.  The WD model has the mass of $M_{\rm WD}= 1.25 ~M_\sun$ with the
mass-accretion rate of $\dot{M}_{\rm acc}= 1\times 10^{-9} ~M_\sun$ yr$^{-1}$.
We set our outburst day to be $t=0=$JD 2460348.0 $=$ UT 2024 February 7.5.
The model $V$ light curve (black line) is calculated from free-free emission
with Equation (\ref{free-free_flux_v-band}) of the nova winds 
\citep{kat26hs} whereas the model X-ray light curve (magenta line) is
calculated with blackbody emission of the WD photosphere (0.3-1.5 keV).
We adopt $\mu_V\equiv (m-M)_V= 15.3$.
\label{v1723_sco_only_model_v_observation_logscale}}
\end{figure*}

GeV gamma-ray and hard X-ray emissions have been detected in some
classical novae.  GeV gamma-ray emissions are observed in an early phase
of a nova and lasts for a few tens of days
\citep[e.g.,][]{abd10, ack14aa, li17mc, gor21ap}.
Hard X-ray emissions are detected in a later intermediate phase
of a nova outburst \citep[e.g.,][]{llo92ob, bal98ko, muk01i}.
Both the GeV gamma-ray and hard X-ray emissions indicate a strong shock,
which is closely related to multiple line systems \citep{ayd20ci,cho21ms}
and dust formation \citep{geh15eh,der17ml,hac25kw}.
If a shocked shell is optically thick, it dominates the optical peak 
luminosity and the nova becomes a superbright nova \citep{hac26kv1674her3}.
In the present paper, we examine how the shock plays a role in V1723 Sco.

The classical nova V1723 Sco was discovered by A. Pearce on UT 2024 February
8.827 (JD 2460349.327) at 7.8 mag \citep{pearce24}.
The outburst evolution is followed by optical, X-ray, and gamma-ray
observations \citep[e.g., ][]{fau26}.  A day after the discovery, 
the Fermi/Large Area Telescope (LAT) detected GeV gamma rays from
V1723 Sco \citep{cheung24}.  \citet{fau26} reported a detailed analysis
of gamma-ray flux from V1723 Sco.
No X-rays were detected with NuSTAR on 1.472 and 2.94 day after the discovery
\citep{sok24a}, but hard X-rays were detected with NuSTAR on 47 day
\citep{fau26}.  The Swift/XRT resulted in no detection of X-rays
from 1.2 day to 13.4 day,
then detected X-rays until $\sim 150$ day.
V1723 Sco entered the SuperSoft X-ray
Source (SSS) phase from 111 day until 150 day \citep{luna24}.

This paper is organized as follows. First we give a quick look at 
light curve fitting of our model with the V1723 Sco observation
and determine the WD mass, mass accretion rate to the WD, and distance
modulus in the $V$ band, $\mu_V\equiv (m-M)_V$, in Section 
\ref{quick_look_light_curve}.
We examine various properties of V1723 Sco associated to the shock formation 
in Section \ref{formation_shock}.  Comparing V1723 Sco with the two
well-studied superbright novae V1500 Cyg and V1674 Her,
we obtain various characteristic
properties of V1723 Sco in Sections \ref{fully_consistent_models}.
Conclusions follow in Section \ref{sec_conclusion}.  
We also obtain $(m-M)_V$ toward V1723 Sco by the time-stretching method
in Appendix \ref{appendix_time-stretching}.

\begin{figure*}
\epsscale{0.95}
\plottwo{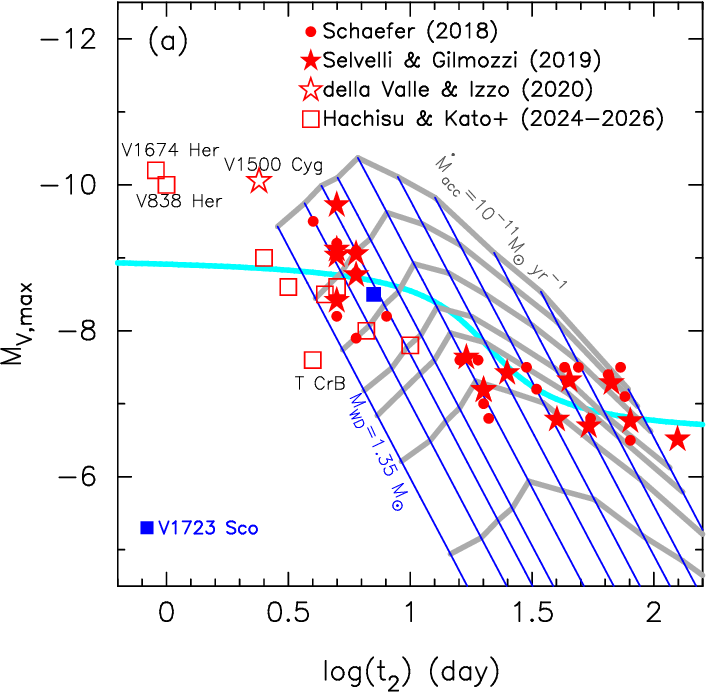}{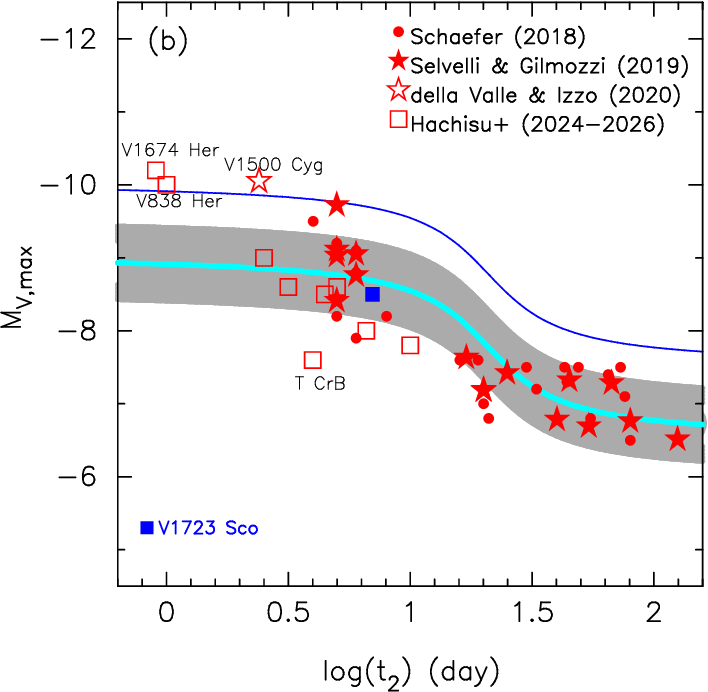}
\caption{
(a) MMRD diagram for novae.  The blue lines denote equi-WD mass line of 
$M_{\rm WD}= 1.35$, 1.3, 1.25, 1.2, 1.1, 1.0, 0.9, 0.8, 0.7, and $0.6~M_\sun$.
The thick solid gray lines depict equi-mass accretion rate line of 
$\dot{M}_{\rm acc}=3\times 10^{-8}$, $1\times 10^{-8}$, 
$5\times 10^{-9}$, $3\times 10^{-9}$, $1\times 10^{-9}$, 
$1\times 10^{-10}$, and $1\times 10^{-11} M_\sun$~yr$^{-1}$.
These blue and gray lines are taken from \citet{hac20skhs}, which are 
calculated from the optically thick nova wind models \citep{kat94h}
and thermonuclear runaway models \citep{hac20skhs}.  The peak brightness of
each nova is calculated from free-free emission with 
Equation (\ref{free-free_flux_v-band}).
We overplot filled red circles taken from ``Golden sample'' of
\citet{schaefer18}, filled stars from \citet{sel19},
and open star (V1500~Cyg) from \citet{del20i}.
The 8 novae (open red squares; KT Eri, V339 Del, V392 Per, V1674 Her,
V838 Her, V597 Pup, V5583 Sgr, and V5589 Sgr) are
taken from \citet{hac25kw}, \citet{hac24km}, \citet{hac25kv392per},
\citet{hac26kv1674her3}, and \citet{hac26ksbright}, respectively.
The thick solid cyan line indicates the empirical line obtained
by \citet{del20i} for an inverse S-shaped MMRD relation.
The position of V1723 Sco (filled blue square) is taken from the present work.
We also add the position of T CrB for comparison.
(b) Same as panel (a), but we show only the position of each nova and
empirical MMRD line of \citet{del20i}.   The thick cyan line
indicates the same as the thick cyan line in panel (a), and light-gray
shadow line corresponds to its $\pm 0.5$ mag region.
The blue line is 1 mag above the thick cyan line.
If the peak absolute $V$ magnitude of a nova is brighter than
the blue line (1 mag brighter than the typical nova brightness),
it is dubbed a superbright nova \citep{del91}.  The three novae, V1500 Cyg,
V838 Her, and V1674 Her are superbright novae \citep{hac26ksbright}.  
\label{max_t2_v1723_sco_selvelli2019_schaefer2018_all_saio_kato2026}}
\end{figure*}

\section{Quick look at light curve fitting}
\label{quick_look_light_curve}

\subsection{Model light curve fitting}
\label{model_fit_light_curve}

Figure \ref{v1723_sco_only_model_v_observation_logscale}
depicts the $V$ light curve of V1723 Sco as well as 
our model $V$ light curve of a 1.25 $M_\sun$ WD taken from \citet{kat26hs}.
Here, we set the origin of time as the outburst day, which is
$t=0=$JD 2460348.0 $=$ UT 2024 February 7.5, that is,
1.327 day before the discovery.
The black line of our model reasonably reproduces the observation
of V1723 Sco until the nebular phase started.
Because strong emission lines such as [\ion{O}{3}] begin to contribute
to the $V$ luminosity in the nebular phase, the black line deviates from
the observation.  This is because our model $V$ light curve of
Equation (\ref{free-free_flux_v-band}) does not include the effect of
line emissions, but represents only continuum flux.

We obtain $\mu_V\equiv (m-M)_V=15.3\pm 0.2$ by direct comparison of
our $V$ light curve with the observed $V$ data. 
We confirm the best-fit value of $(m-M)_V=15.3\pm 0.2$
when we increase $(m-M)_V$ by a 0.1 mag step from $(m-M)_V=$14.3 to 16.3
in Figure \ref{v1723_sco_only_model_v_observation_logscale}
and, in each step, directly fit our model $V$ light curve
(black line) with the observation.
In Appendix \ref{appendix_time-stretching}, we also obtain the same
$(m-M)_V=15.3\pm 0.2$, in a different manner,
without model light curves, by the time-stretching method.

We add the Swift/XRT count rate light curves taken from the Swift website
\citep{eva09}, for the both soft component
(cyan asterisk: 0.3-1.5 keV) and hard component (gray plus: 1.5-10 keV). 
Also, we add our theoretical 0.3-1.5 keV light curve (magenta line)
calculated from the same nova model for the optical $V$ light curve. 
We are able to broadly reproduce the soft X-ray component light curve
with our blackbody model flux from the WD photosphere.  For the hard X-ray
(1.5-10 keV), we assign them to the shock origin in Section
\ref{formation_shock}, which arises far outside the WD photosphere.

We may conclude that (1) our model $V$ light curve of a $1.25 ~M_\sun$
WD with $\dot{M}_{\rm acc}=1\times 10^{-9} ~M_\sun$ yr$^{-1}$
reproduces the V1723 Sco $V$ light curve, (2)
the same model also broadly explains the behavior of the soft X-ray
(0.3-1.5 keV) light curve, and (3) the first GeV gamma-ray detection
with the Fermi/LAT is almost coincident with the peak brightness of
our model $V$ light curve. 

In the present paper,
we use \citet{kat26hs}'s fully self-consistent nova explosion models to
analyze the multiwavelength light curves of V1723 Sco.  
For a given set of the WD mass
and mass-accretion rate to the WD, they calculated the WD structures
from the center of the WD up to the WD photosphere, based on their own
Henyey type evolution code consistently combined with steady state wind
mass loss solutions as a surface boundary condition.

Because early nova spectra can be approximated by free-free emission
\citep[e.g., ][]{gal76,enn77}, we calculate the $V$ brightness
of \citet{kat26hs}'s nova model by free-free emission luminosity
as \citep{hac06kb, hac20skhs}
\begin{equation}
L_{V, \rm ff,wind} = A_{\rm ff} ~{{\dot M^2_{\rm wind}}
\over{v^2_{\rm ph} R_{\rm ph}}},
\label{free-free_flux_v-band}
\end{equation}
where $\dot{M}_{\rm wind}$ is the wind mass loss rate, 
$v_{\rm ph}$ the velocity at the photosphere, 
and $R_{\rm ph}$ the photospheric radius  
(see Equation (3) in \citet{kat25hs} for details on 
the coefficient $A_{\rm ff}$ and how to determine it for a specific nova).
The wind mass loss rate reaches maximum at the maximum expansion of
the WD photosphere.  Therefore, the $V$ luminosity reaches maximum at the
maximum expansion of the WD photosphere.

Free-free emission comes from optically thin plasma outside
the WD photosphere and dominates the $V$ luminosity.
Therefore, the $V$ luminosity is not constrained by the Eddington limit,
which works in the optically thick region.

These model $V$ light curves are already introduced
in \citet{kat25hs, kat26hs} and \citet{hac25kv1674her2,
hac26kv1674her3} for model light curve fittings with V1674 Her, in
\citet{kat26hs} for four novae of KT Eri, V339 Del, V597 Pup,
and SMC Nova 2016-10a, in \citet{hac26ksbright} for eight novae of
V1500 Cyg, V1674 Her, CP Lac, CP Pup,
V838 Her, V597 Pup, V5583 Sgr, and V5589 Sgr.

\begin{figure}
\epsscale{1.15}
\plotone{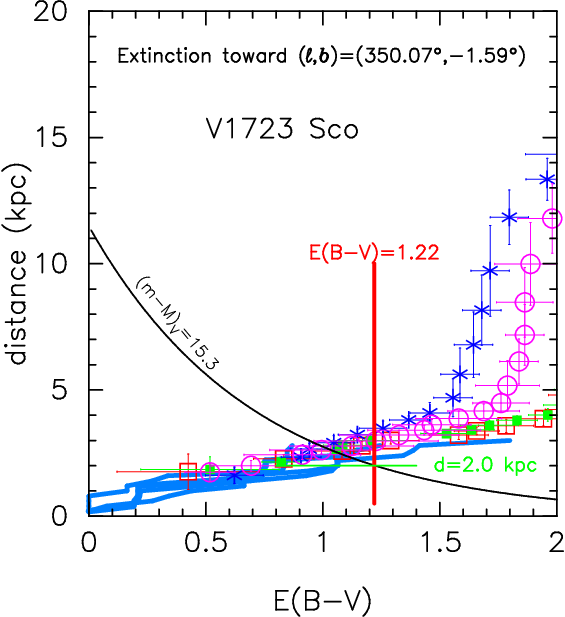}
\caption{
The distance-reddening relations toward the galactic coordinates of
$(\ell, b)= (350\fdg07, -1\fdg59)$, that is, toward V1723 Sco.
The black line denotes Equation (\ref{distance_reddening_law})
with $(m-M)_V=15.3$.  We add four thick cyan-blue lines of 
\citet{chen19}, which are nearby directions of 
$(\ell, b)= (350\fdg05, -1\fdg65)$, $(350\fdg05, -1\fdg55)$,
$(350\fdg15, -1\fdg65)$, and $(350\fdg15, -1\fdg55)$.
The unfilled red squares, filled green squares, blue asterisks,
and unfilled magenta circles, each with error bars,
represent the relation given by \citet{mar06}.
The black line crosses the green line at $E(B-V)=1.22$.
}\label{distance_reddening_v1500_cyg_xxx_no2}
\end{figure}

\subsection{Position in the MMRD diagram}
\label{position_mmrd_diagram}

The maximum magnitude versus rate of decline (MMRD) diagram has been used
to discuss nova properties \citep[e.g., ][]{del20i}.  \citet{hac20skhs}
calculated a number of nova model light curves with Equation 
(\ref{free-free_flux_v-band}), and constructed theoretical MMRD diagram.
Figure \ref{max_t2_v1723_sco_selvelli2019_schaefer2018_all_saio_kato2026}(a)
shows such an MMRD diagram for various models of classical novae.
The gray and blue lines indicate the equi-$\dot{M}_{\rm acc}$ and
equi-$M_{\rm WD}$ lines, respectively, of our database of theoretical
light curve models for various sets of WD mass ($M_{\rm WD}$)
and mass-accretion rate ($\dot{M}_{\rm acc}$).
For comparison, we also plot a popular MMRD diagram in Figure 
\ref{max_t2_v1723_sco_selvelli2019_schaefer2018_all_saio_kato2026}(b).

We have measured the $t_2$ time of V1723 Sco to be $t_2=7$ day from
the $V$ data of \citet[magenta circles,][]{habtie24} and
the American Association of Variable Star Observers (AAVSO,
filled green squares) as depicted in Figure 
\ref{v1723_sco_only_model_v_observation_logscale}.
The maximum absolute $V$ magnitude of V1723 Sco is estimated to be
$M_{V,\rm max}= m_{V,\rm max} -(m-M)_V = 6.77 - 15.3 = -8.5$.
Here, we adopt $m_{V,\rm max}\equiv V_{\rm max}=6.77$ (filled magenta circle
in Figure \ref{v1723_sco_only_model_v_observation_logscale})
from \citet{habtie24} and $(m-M)_V=15.3$ obtained in Section 
\ref{model_fit_light_curve} above. 

We plot the MMRD point of V1723 Sco (filled blue square) in Figure
\ref{max_t2_v1723_sco_selvelli2019_schaefer2018_all_saio_kato2026}.
V1723 Sco is located closely on the typical inverse S-shaped MMRD relation
(thick cyan line) given by \citet{del20i}, 
so that V1723 Sco is not a superbright nova
defined by \citet{del91}, but a normal very fast \citep[$t_2\le 10$ day;
][]{pay57} nova.  We discuss the difference in the properties between
normal novae and superbright novae in Section \ref{fully_consistent_models}. 

The position of V1723 Sco is close to the blue line of 
$M_{\rm WD}=1.25 ~M_\sun$ and to the thick gray line of
$\dot{M}_{\rm acc}=1\times 10^{-9} ~M_\sun$ yr$^{-1}$ in Figure 
\ref{max_t2_v1723_sco_selvelli2019_schaefer2018_all_saio_kato2026}(a).  
This position is roughly consistent with our best fit model for V1723 Sco,
that is, $M_{\rm WD}=1.25 ~M_\sun$ with $\dot{M}_{\rm acc}=1\times 10^{-9} 
~M_\sun$ yr$^{-1}$ in Figure 
\ref{v1723_sco_only_model_v_observation_logscale}.

\subsection{Distance and Reddening}
\label{sec_distance}


We estimate the distance, $d$, and reddening, $E(B-V)$, toward V1723 Sco.
Figure \ref{distance_reddening_v1500_cyg_xxx_no2} depicts various
distance-reddening relations toward V1723 Sco.
The black line indicates the relation of
\begin{equation} 
(m-M)_V=3.1 E(B-V) + 5 \log (d/{\rm 10 ~pc})
\label{distance_reddening_law}
\end{equation}
together with $(m-M)_V=15.3$. 

The Gaia DR3 parallax of V1723 Sco, 
$\varpi= (1\farcs 0915607 \pm 0\farcs 40312746)\times 10^{-3}$
for the source ID 5974053153711274112,
gives a distance of $d_{\rm geometic}= 1.735$ (1.085--3.160) kpc
or $d_{\rm photogeometic}= 2.016$ (1.721--3.107) kpc \citep{bai21rf}.
Here, we adopt $d=2.0^{+1.1}_{-0.3}$ kpc (photogeometric) for V1723 Sco.
The black line of $(m-M)_V=15.3$ crosses the green line of $d=2.0$ kpc
at the reddening of $E(B-V)=1.22$.  Thus, we obtained the reddening
of $E(B-V)=1.22$ toward V1723 Sco. 

\citet{fau26} obtained the reddening of $E(B-V)= A_V/3.1 = (3.8\pm 0.16)/3.1
= 1.22\pm0.05$ from the hydrogen column density of 
$N_{\rm H}=8.5\times 10^{21}$ cm$^{-2}$ \citep{sok24b} together with
the $N_{\rm H}$-$A_V$ relation given by \citet{guv09}, where $A_V$ is
the absorption in the $V$ band. 
They further estimated the peak absolute $V$ magnitude of
$M_{V,\rm max}=-8.3$ from an empirical MMRD relation \citep[see Equation
(1) of ][]{fau26}, and obtained the distance of $d=1.9\pm1.1$ kpc from
$(m-M)_V= m_{V,\rm max}-M_{V,\rm max}= 6.77-(-8.3)=15.1$ together with
Equation (\ref{distance_reddening_law}) and $E(B-V)=1.22$.  Their results are
broadly consistent with our results of $d=2.0$ kpc and $E(B-V)=1.22$.
This is because their empirical MMRD relation accidentally gives
a correct (or approximate) brightness for V1723 Sco (see Figure
\ref{max_t2_v1723_sco_selvelli2019_schaefer2018_all_saio_kato2026}).


\begin{figure*}
\epsscale{0.7}
\plotone{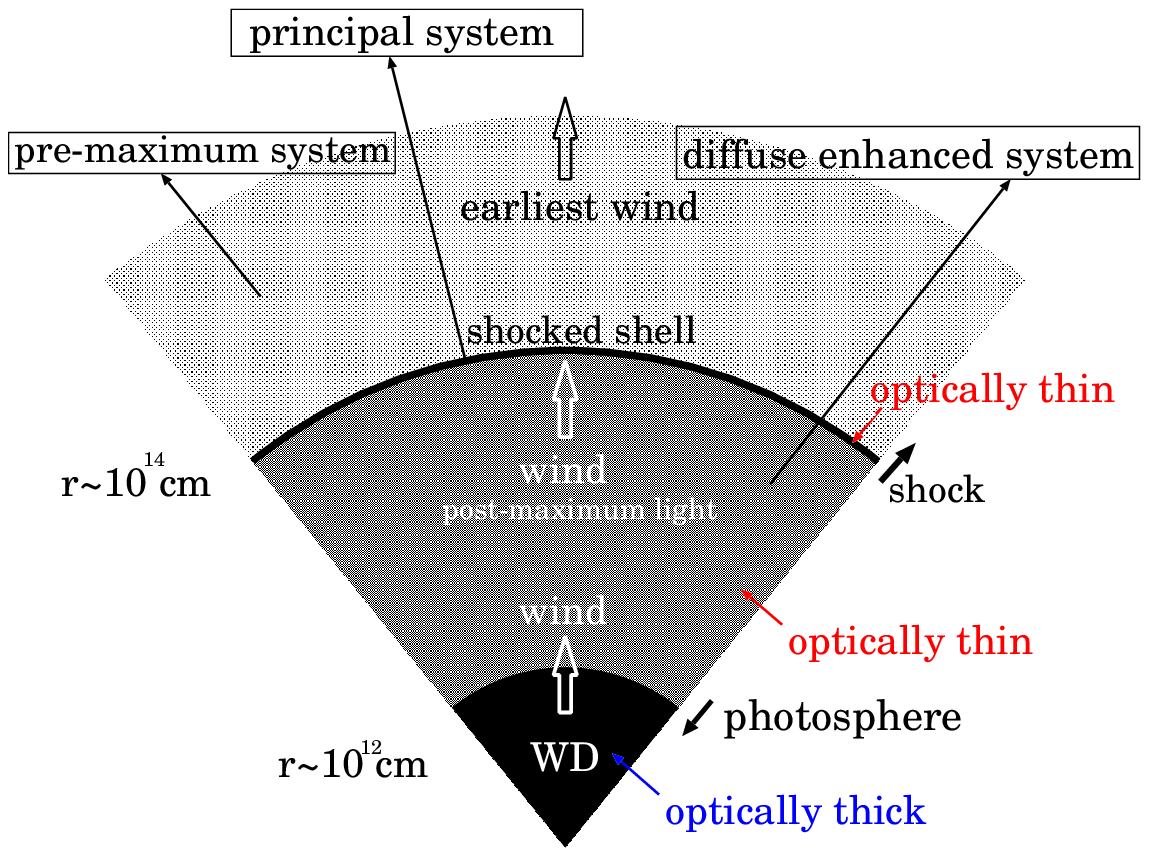}
\caption{
Cartoon for the V1723 Sco shock model.
The nova (WD) photosphere becomes larger than $R_{\rm ph} \sim 0.1 ~R_\sun$
and optically thick winds emerge from the photosphere
\citep{kat22sha, kat25hs}.
After the maximum expansion of the WD photosphere (typically, a few to 
several tens $R_\sun$), it turns to shrink.
A shock forms far outside the WD photosphere \citep{hac22k} as shown
in the cartoon.  This shocked shell is expanding and the optical depth
of the shocked shell is typically less than unity, that is, optically thin.
Then the ejecta is divided into three parts, outermost part of
earliest wind, shocked shell, and inner wind.  
Free-free emission from the nova wind dominates the nova spectrum and
each part contributes to pre-maximum, principal,
and diffuse enhanced absorption/emission line systems \citep{mcl42},
respectively, as proposed by \citet{hac22k, hac23k}.
The velocity of principal system is typically about a half of that of
diffuse enhanced system \citep{mcl42, hac22k}.
\label{wind_shock_config}}
\end{figure*}

\section{Formation of a shock}
\label{formation_shock}

It has been suggested that GeV gamma-ray and hard X-ray emissions
originate from a shock between shells ejected with different velocities
\citep{cho14ly, met15fv, mar18dj}.
If an inner shell is ejected with a larger velocity than that of
an outer shell, the inner shell can collide with the
outer shell and this collision forms a shock 
\citep[e.g.,][]{muk19s, ayd20ci, ayd20sc}.

However, no theoretical explanation has been presented that
naturally explains all these different wavelength observations
based on nova explosion models \citep[for a recent review, see][]{cho21ms}.
Many numerical calculations have been presented from the early
thermonuclear runaway to the extended phase of nova outbursts 
\citep[e.g.,][]{pri92k, pri95k, epels07, sta09ih, den13hb, chen19,
kat22sha, kat22shb}.
These works clearly showed that mass ejection itself is continuous,
no shock forms, and no multiple mass ejection occurs.

\citet{hac22k} showed that a strong shock forms far outside the WD
photosphere, as illustrated in Figure \ref{wind_shock_config},
based on \citet{kat22sha}'s fully self-consistent nova explosion model.  
This is because the velocity of nova ejecta at the WD photosphere
continuously and smoothly increases with time after the maximum expansion
of the WD photosphere \citep[see, e.g., Figure 1 of ][]{hac22k}.
The later ejected matter has a larger expansion velocity so that
it catches up with the former ejected matter and makes a strong shock.
Thus, a shock is formed after the maximum expansion of the WD photosphere
($=$optical $V$ maximum) and propagates far outside the WD photosphere. 
This mechanism of shock formation reasonably explains gamma-ray emission 
and hard X-ray detection/nondetection in classical novae \citep[e.g.,
YZ Ret, V339 Del, and V392 Per in][respectively]{hac23k, hac24km,
hac25kv392per}. 

In many novae, shocked shell is optically thin as illustrated in Figure
\ref{wind_shock_config}, so that the $V$ luminosity is dominated by
free-free emission calculated by Equation (\ref{free-free_flux_v-band}).  
On the other hand, in a few novae, their shocked shells are optically thick
in the very early phase near their optical maxima.  In this case,
the $V$ luminosity is determined by the photosphere of the shocked shell,
the spectrum of which is not free-free emission but approximated
by blackbody \citep[e.g.,][and see Section \ref{full_v1500_cyg_1975}
for more detail on V1500 Cyg]{gal76,enn77}.  These optically thick shocks
are similar to a recombination front of hydrogen in Type II Plateau (IIP)
supernovae \citep[e.g.,][]{dub25}. 
\citet{hac26kv1674her3} regarded that such an optically thick shocked shell
nova is a superbright nova defined by \citet{del91}. 
We discuss such examples in more detail in Section 
\ref{fully_consistent_models} and show that the shocked shell of V1723 Sco
is optically thin.

\subsection{GeV gamma-ray detection}
\label{gev_gamma-ray_detection}

\citet{fau26} analyzed the Fermi/LAT observation and obtained 
the first GeV gamma-ray detection on day 2.0-2.25
after the outburst in our 
assumption of $t_{\rm OB}=$ UT 2024 February 7.5 as shown in Figure
\ref{v1723_sco_only_model_v_observation_logscale}.
Our 1.25 $M_\sun$ WD outburst model with $\dot{M}_{\rm acc}=1\times 
10^{-9} ~M_\sun$ yr$^{-1}$ (black line in Figure 
\ref{v1723_sco_only_model_v_observation_logscale})
reaches its optical $V$ maximum on day 2.0--2.5.
Therefore, the first gamma-ray detection is almost coincident with
the formation of a strong shock based on \citet{hac22k}'s nova shock
model mentioned above.
We estimate the thermal shock energy based on our 1.25 $M_\sun$ WD model
in Section \ref{shock_gamma-ray_flux} below, 
which is consistent with the GeV gamma-ray flux obtained by \citet{fau26}.

\subsection{Spectral evolution}
\label{sepctral_evolution}

Figure \ref{wind_shock_config} illustrates our shock model in the ejecta of
V1723 Sco: the shocked shell divides a nova ejecta into three parts
with different velocities.  \citet{hac22k} identified the pre-maximum,
principal, and diffuse enhanced absorption/emission line systems
proposed by \citet{mcl42} as the velocities of the earliest wind,
shocked shell, and inner wind, respectively.
Therefore, we are able to determine the velocity of each part
from nova spectral features. 

\citet{ayd24} reported an optical spectrum of V1723 Sco
on UT 2024 February 9.84 ($=$ day 2.34)
at an unfiltered visible magnitude of 7.3 mag.
Their low- and medium-resolution spectra of the 4.1 m SOAR telescope
show absorption lines and P Cygni profiles of Balmer, \ion{Fe}{2},
\ion{Na}{1}, and \ion{O}{1}.  The absorption trough at H$\alpha$
is at a blue-shifted velocity of around 1200 km s$^{-1}$.
The spectrum is that of a classical nova near optical peak.
We regard this absorption trough as the principal absorption/emission line
system proposed by \citet{mcl42}, which appeared near optical maximum
(see Figure \ref{wind_shock_config}).

\citet{habtie24} reported spectroscopy on UT 2024 February 11.368 and
12.362 ($=$ day 3.869 and 4.862) and concluded that, from H$\alpha$ and
H$\beta$ lines, shell is expanding at a velocity of
around 3200$\pm200$ and 1800$\pm100$ km s$^{-1}$, respectively.

\citet{sho24} also reported the spectroscopy on UT 2024 February 11.3 
($=$ day 3.8), February 12.3 (day 4.8), February 13.3 (day 5.8).
\ion{Na}{1} D with a broad absorption extending to about $-3000$ km s$^{-1}$
(day 3.8) with a maximum depth at $-1100$ km s$^{-1}$ (day 3.8)
and $-1600$ km s$^{-1}$ (day 4.8).
The Balmer lines all show strong, very asymmetric profiles that developed
rapidly over the three days. The H$\alpha$ emission wing extended to
$-3000$ km s$^{-1}$ on the first two days but the profile blue
emission edge was at $-1200$ km s$^{-1}$ (day 3.8);
it extended to $-3500$ km s$^{-1}$ on day 4.8 with the asymmetry
appearing more like a nascent absorption.

We regard the broad absorption component as the diffuse enhanced 
absorption/emission line system proposed by \citet{mcl42}
(see Figure \ref{wind_shock_config}).

To summarize, (1) the principal absorption line system appeared on day
2.3 near the $V$ maximum ($=$ day 2.0--2.5) and its expansion velocity
is $v_{\rm p}\sim 1200$ km s$^{-1}$ while (2) the diffuse enhanced
absorption line system arose a few days after the optical $V$ maximum
and its largest expansion velocity is $v_{\rm wind}= v_{\rm d}\sim
3000$ km s$^{-1}$.  The configuration of nova winds and shocked shell
is illustrated in Figure \ref{wind_shock_config}.

\subsection{Temperature behind the shock}
\label{temperature_shock}

In the previous subsection,
we regard the velocity $v_{\rm shock}$ of
this shocked shell to be the principal absorption system of V1723 Sco,
$v_{\rm shock}\approx v_{\rm p}= 1200$ km s$^{-1}$ and the velocity of inner
winds emerging from the WD photosphere to be the diffuse enhanced
absorption system, $v_{\rm wind}\approx v_{\rm d}\sim 3000$ km s$^{-1}$.

The temperature of shocked matter and shock energy can be estimated from
the difference between the fresh ejecta (inner wind) velocity
and expanding speed of the shocked shell, i.e.,
\begin{eqnarray}
kT_{\rm sh}& \sim & {3 \over 16} \mu m_p
\left( v_{\rm wind} - v_{\rm shock} \right)^2 \cr
& \approx & 1.0 {\rm ~keV~}
\left( {{v_{\rm wind} - v_{\rm shock}} \over
{1000 {\rm ~km~s}^{-1}}} \right)^2,
\label{shock_kev_energy}
\end{eqnarray}
where $k$ is the Boltzmann constant, $T_{\rm sh}$ is the temperature
at the shock \citep[see, e.g.,][]{met14hv},
$\mu$ is the mean molecular weight ($\mu =0.5$ for hydrogen plasma),
and $m_p$ is the proton mass.
Substituting $v_{\rm shock}= v_{\rm p}=1200$ km s$^{-1}$ and
$v_{\rm wind}= v_{\rm d}=3000$ km s$^{-1}$,
we obtain the post-shock temperature
$k T_{\rm sh}\sim 3.3$ keV.
This shock temperature is consistent with the thermal plasma temperature of
the NuSTAR observation on day 48, that is,
$k T \approx 3.7$ keV obtained by \citet{fau26}.

\subsection{Luminosity of GeV gamma-ray from a shock}
\label{shock_gamma-ray_flux}

The shock properties constrain
\begin{equation}
L_\gamma = \epsilon_{\rm nth} \epsilon_\gamma L_{\rm sh} \sim 10^{35} -
10^{36} {\rm ~erg~s}^{-1},
\label{observed_gamma_ray_flux}
\end{equation}
from observation \citep{ack14aa}.
Here, $L_{\gamma}$ is the gamma-ray
luminosity between 100 MeV and 300 GeV (Fermi/LAT window),
$\epsilon_{\rm nth} \lesssim 0.1$ is the fraction of the shocked
thermal energy ($L_{\rm sh}$) used to accelerate
relativistic nonthermal particles,
$\epsilon_\gamma \lesssim 0.1$ is
the fraction of this energy radiated in the Fermi/LAT band
\citep[typically $\epsilon_{\rm nth} \epsilon_\gamma
\lesssim 0.03$;][]{met15fv}.
Equation (\ref{observed_gamma_ray_flux}) and
$\epsilon_{\rm nth} \epsilon_\gamma
\sim 0.01$ simply requires
\begin{equation}
L_{\rm sh} \sim 10^{37} - 10^{38} {\rm ~erg~s}^{-1}.
\label{shocked_generation_energy_flux}
\end{equation}

The shock thermal energy at a reverse shock is estimated to be
\citep{met14hv}
\begin{eqnarray}
L_{\rm sh}& \sim & {{9}\over {32}} {\dot M}_{\rm wind}
{{( v_{\rm wind} - v_{\rm shock} )^3} \over {v_{\rm wind}}} \cr
&=& 1.8\times 10^{37}{\rm ~erg~s}^{-1}
\left( {{{\dot M}_{\rm wind}} \over
{10^{-4} ~M_\sun {\rm ~yr}^{-1}}} \right) \cr
 &  & \times
\left( {{{v_{\rm wind} - v_{\rm shock}} \over {1000{\rm ~km~s}^{-1}}}}
\right)^3
\left( {{{1000{\rm ~km~s}^{-1}} \over {v_{\rm wind}}} }\right).
\label{shocked_energy_generation}
\end{eqnarray}
Substituting $\dot{M}_{\rm wind}= 1.4 \times 10^{-4} ~M_\sun$ yr$^{-1}$
from our model ($1.25 ~M_\sun$ WD with $\dot{M}_{\rm acc}=1\times
10^{-9} ~M_\sun$ yr$^{-1}$),
$v_{\rm shock}= v_{\rm p}=1200$ km s$^{-1}$, and
$v_{\rm wind}= v_{\rm d}=3000$ km s$^{-1}$,
we obtain the post-shock energy flux of $L_{\rm sh} \sim 5\times 10^{37}$
erg s$^{-1}$ just after maximum (on day 2.0).
This energy flux decreases to 
$L_{\rm sh} \sim 1\times 10^{37}$ erg s$^{-1}$ on day 11.
This decline rate is broadly consistent with the gamma-ray luminosity
light curve in Figure 1 of \citet{fau26} if $L_{\gamma} \propto L_{\rm sh}$.

The optical luminosity at maximum (on day 2.0) reaches $L_{\rm opt}\sim 
7\times 10^{38}$ erg s$^{-1}$ in our model.  Then we have the ratio of 
$L_{\rm sh}/L_{\rm opt}= 5\times 10^{37}/ 7\times 10^{38} \sim 0.1$
at the optical maximum.
\citet{fau26} estimated the ratio of $L_{\gamma}/L_{\rm opt}= 
10^{-3.2}$--$10^{-2.5}$.  Therefore, the efficiency of $\epsilon_{\rm nth} 
\epsilon_\gamma = L_{\gamma}/L_{\rm sh}=(L_{\gamma}/L_{\rm opt})/
(L_{\rm sh}/L_{\rm opt}) =(10^{-3.2}$--$10^{-2.5})/0.1=
10^{-2.2}$--$10^{-1.5} \lesssim 0.03$ is satisfied
with the requirement of $\epsilon_{\rm nth} \epsilon_\gamma < 0.03$
in \citet{met15fv}.

\begin{figure}
\epsscale{1.0}
\plotone{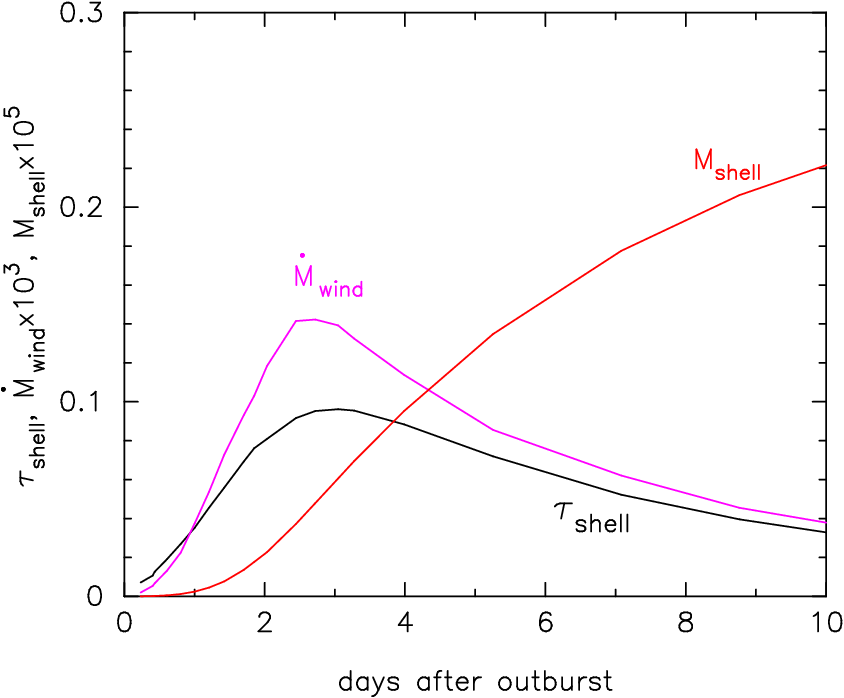}
\caption{The optical depth $\tau_{\rm shell}$ 
of the ejected shell is plotted against the time, days after outburst,
for our 1.25 $M_\sun$ WD model with
$\dot{M}=1\times 10^{-9} ~M_\sun$ yr$^{-1}$.
It is as small as $\tau_{\rm shell}\lesssim 0.1$, which clearly indicates 
that the shell is optically thin.
We also plot the evolutions of wind mass-loss rate $\dot{M}_{\rm wind}$
in units of $M_\sun$ yr$^{-1}$ and ejected shell mass $M_{\rm shell}$
in units of $M_\sun$.  
\label{v1723_sco_optical_depth_linear}}
\end{figure}

\subsection{Optical depth of the shell}
\label{optical_depth_shock}

Here, we estimate the optical depth of the shell.
The optical depth $\tau_{\rm shell}$ is approximately calculated from
\begin{equation}
\tau_{\rm shell} \equiv \int_{\rm shell} \kappa \rho d r
 \approx  {\kappa {M_{\rm shell}} \over {4\pi R^2_{\rm shell}}},
\label{potical_depth_shell}
\end{equation}
where $\kappa$ is the opacity, $\rho$ the density, $r$ the radius from
the center of the WD, and $M_{\rm shell}$ the mass of, and
$R_{\rm shell}$ the radius of, the shell.

We take the opacity of $\kappa\sim 1$ g$^{-1}$ cm$^2$.
The radius of the shocked shell is approximated by
$R_{\rm shell}\approx t\times v_{\rm shell}\sim t\times v_{\rm p}$.
More exactly, we use the expansion velocity of the shell of
$v_{\rm shell}=v_{\rm p}= 1200$ km s$^{-1}$.
We plot the calculated optical depth $\tau_{\rm shell}$ in Figure
\ref{v1723_sco_optical_depth_linear}.
The optical depth is as small as $\tau_{\rm shell}\lesssim 0.1$,
which clearly indicates that the shell is optically thin.

The mass in the earliest wind in Figure \ref{wind_shock_config} is calculated
to be $M_{\rm ew}= \int \dot{M}_{\rm wind} d t \approx 3\times 10^{-7} 
~M_\sun$ at the optical maximum ($=$at the maximum wind mass loss rate).
Then, it gradually decreases because the shocked shell absorbs a part of
the earliest wind.  The mass of the shocked shell increases with time
after the  optical maximum ($=$after the maximum wind mass loss rate),
that is, $M_{\rm sh}= \int \dot{M}_{\rm wind} d t$.  It soon becomes
$\sim 3\times 10^{-7} ~M_\sun$ a few days after a shock arises ($=$a few days
after the optical maximum) and finally to $\sim 3\times 10^{-6} ~M_\sun$.
The mass of the inner wind is as small as $M_{\rm iw} \lesssim
1\times 10^{-7} ~M_\sun$ a few days after optical maximum. 
This is because the wind mass loss rate is extremely high 
($\dot{M}_{\rm wind}\sim 1.4 \times 10^{-4} ~M_\sun$ yr$^{-1}$)
at/near the optical maximum.

\subsection{Shock duration}
\label{shock_duration}

Hard X-rays are emitted from the hot plasma behind the shock.
The hard X-ray flux could substantially decay after the shock disappears. 
The shock is alive even after the winds stop because it takes a time
that the latest wind reaches the shock.  This elapse time is 
estimated by \citet{hac23k} and its total shock duration 
$\tau_{\rm shock}$ becomes
\begin{equation}
\tau_{\rm shock}= {{t_{\rm ws}} \over
{\left( 1- {{v_{\rm p}} \over {v_{\rm d}}}\right)}},
\label{duration_of_shock}
\end{equation}
where $t_{\rm ws}$ is the wind stopping time from the epoch of shock 
formation.
The nova winds of our 1.25 $M_\sun$ WD model stop on day 85, as shown in 
Figure \ref{v1723_sco_only_model_v_observation_logscale}. 
Substituting $v_{\rm shock}\approx v_{\rm p}\sim 1200$ km s$^{-1}$
(principal system), $v_{\rm wind}\approx v_{\rm d}\sim 3000$ km s$^{-1}$
(diffuse enhanced system), and $t_{\rm ws}=85 - 2=83$ day 
(the wind duration since the shock arises) into Equation 
(\ref{duration_of_shock}),
we obtain the shock duration of $\tau_{\rm shock}= 83/0.6= 138$ days.
Therefore, in V1723 Sco, we expect hard X-ray emission
until day (138+2=) 140.  
This time corresponds broadly to the epoch of rapidly decay of
the hard X-ray component (gray plus symbols
in Figure \ref{v1723_sco_only_model_v_observation_logscale}).

\subsection{Hydrogen column density of a shocked shell}
\label{column_density_shocked_shell}

The hydrogen column density is estimated from
$M_{\rm shell}= 4 \pi R_{\rm sh}^2 \rho h_{\rm shell}$,
where $M_{\rm shell}$, $R_{\rm sh}$, $\rho$, and $h_{\rm shell}$ are 
the mass, radius, density, and the thickness of the shocked shell.
If we take an averaged velocity of shell $v_{\rm sh}=
v_{\rm shell}= v_{\rm shock}= 1200$ km s$^{-1}$,
the shock radius is calculated from $R_{\rm sh}(t)= v_{\rm shock}\times t$.
This reads
\begin{eqnarray}
N_{\rm H} & = & {{X \over m_p} {{ M_{\rm shell} }
\over {4 \pi R^2_{\rm sh}}}} \cr
 & \approx & 4.8\times 10^{21} {\rm ~cm}^{-2}
\left({X \over {0.5}}\right)
\left( {{M_{\rm shell}} \over {10^{-6} M_\sun}} \right)
\left( {{R_{\rm sh}} \over {10^{14} {\rm ~cm}}} \right)^{-2}
\cr
 & \approx & 6.4 \times 10^{21} {\rm ~cm}^{-2}
\left({X \over {0.5}}\right)
\left( {{M_{\rm shell}} \over {10^{-6} M_\sun}} \right) \cr
& & \times
\left( {{v_{\rm shell}} \over {1000 {\rm ~km~s}^{-1}}} \right)^{-2}
\left( {{t} \over {10~{\rm day}}} \right)^{-2}.
\label{column_density_hydrogen_time}
\end{eqnarray}
Thus, the column density decreases from 
$N_{\rm H} \approx 2\times 10^{23}$ cm$^{-2}$ on day $t=3$, to 
$\approx 1.5\times 10^{22}$ cm$^{-2}$ on day $t=11$, to
$\approx 4\times 10^{21}$ cm$^{-2}$ on day $t=22$,
and to $\approx 9\times 10^{20}$ cm$^{-2}$ on day $t=44$ for 
the shell mass of $M_{\rm shell}=3 \times 10^{-6}~M_\odot$ 
in our model (1.25 $M_\sun$ WD with $\dot{M}_{\rm acc}=1\times 10^{-9}
M_\sun$ yr$^{-1}$).
This $N_{\rm H}\approx 4\times 10^{21}$ cm$^{-2}$ of the shell (on day $t=22$)
is comparable to the interstellar hydrogen column density of
$N_{\rm H}\sim 8.5\times 10^{21}$ cm$^{-2}$ \citep{fau26}.

The hard X-ray component (1.5-10.0 keV) count rate 
is shown by the gray plus symbols in Figure
\ref{v1723_sco_only_model_v_observation_logscale}.
It increases from day 15 and reaches maximum
on day 30.  This increase is owing to the rapid decay of hydrogen
column density of the ejecta as calculated above.
Then, it starts to decay from day 100,
followed by a rapid decay from day 140.  The hard X-ray
component is detected even until day $\sim 150$, the end of observation.
This sharp drop in the hard X-ray flux on day $t \lesssim 140$ day
is broadly consistent with our estimate of the shock duration until
day $\sim 140$, estimated in Section \ref{shock_duration}.



\begin{figure*}
\gridline{\fig{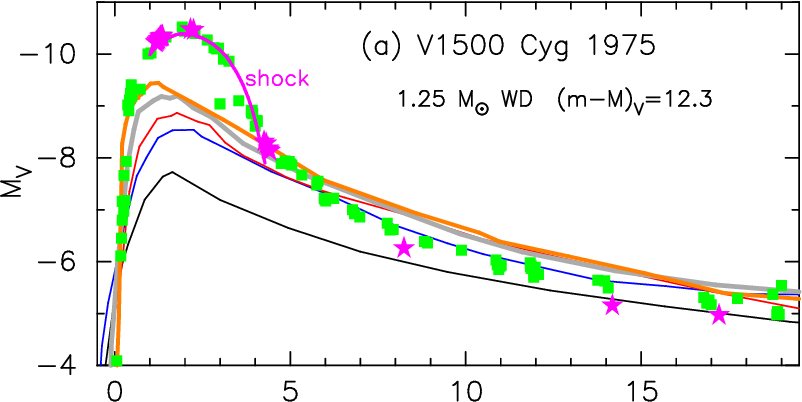}{0.5\textwidth}{}
          }
\gridline{
          \fig{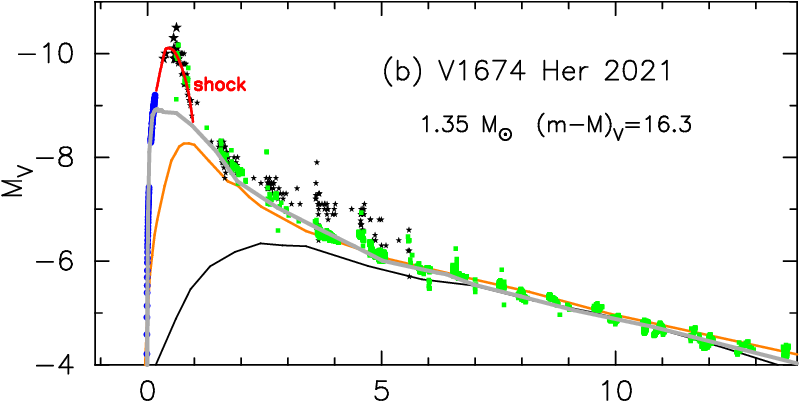}{0.5\textwidth}{}
          }
\gridline{
          \fig{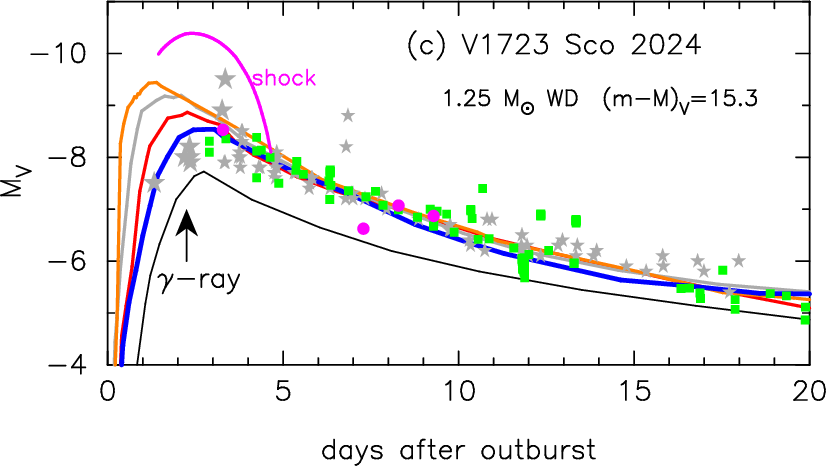}{0.5\textwidth}{}
          }
\caption{
(a) The $V$ (filled green squares) and Str\"omgren $y$ (filled magenta stars)
light curves of V1500 Cyg on a linear timescale.  The $V$/$y$ data are
the same as those in Figure 7(a) of \citet{hac26kv1674her3}.
We add five free-free emission model $V$ light curves of
a 1.25 $M_\sun$ WD with $\dot{M}_{\rm acc}= 
5\times 10^{-11}$ (thick orange line),
$1\times 10^{-10}$ (gray line),
$5\times 10^{-10}$ (red line),
$1\times 10^{-9}$ (blue line),
and $5\times 10^{-9}$ (black line)
$M_\sun$ yr$^{-1}$.
The thick magenta line labeled shock indicates the brightness of
the optically thick shocked shell for V1500 Cyg, which is
taken from \citet{hac26kv1674her3}.
(b) Same as in panel (a), but for the $V$, $g$, and visual light
curves of V1674 Her.  The $V$, $g$, and visual data are the same
as those in Figure 1 of \citet{hac26kv1674her3}.
We add three model $V$ light curves of a 1.35 $M_\sun$ WD with
$\dot{M}_{\rm acc}= 
1\times 10^{-11}$ (thick gray line),
$5\times 10^{-10}$ (orange line), and $5\times 10^{-9}$ (black line)
$M_\sun$ yr$^{-1}$.  The thick red line labeled shock indicates
the brightness of the optically thick shocked shell for V1674 Her,
which is taken from \citet{hac26kv1674her3}.
(c) Same as in panel (a), but for the $V$, visual, and unfiltered $V$
($=$ clear $V$ $=$ CV) light curves of V1723 Sco.
We assume the outburst day of $t_{\rm OB}=$ JD 2460348.0.
The first detection day of GeV gamma-rays is denoted by the
upward arrow labeled $\gamma$-ray, which broadly coincides with the $V$ peak
(blue line) of $\dot{M}_{\rm acc}=1\times 10^{-9} ~M_\sun$ yr$^{-1}$.
The filled green squares and magenta circles denote the $V$ data from
AAVSO/VSOLJ and \citet{habtie24}, respectively.
The filled gray stars are visual or
unfiltered $V$ taken from \citet{pearce24} and AAVSO.
See the main text for more details.
\label{v1500_cyg_v1674_her_v1723_sco_full_model_fit_linear}}
\end{figure*}

\section{Fully self-consistent nova explosion models}
\label{fully_consistent_models}

In this section, we discuss the properties of V1723 Sco light curve 
by comparing it with V1500 Cyg and V1674 Her light curves, both of which
are superbright novae.
This is partly because the peak brightness of V1723 Sco becomes 
$M_{V, \rm max}= -9.5$ and is 1 mag brighter than $M_{V, \rm max}= -8.5$
in Section \ref{position_mmrd_diagram}
if we take serious the two brightest unfiltered
$V$ magnitudes of $m_v=6.4$ (day 3.26) and $m_v=5.8$ (day 3.35) in Figure
\ref{v1723_sco_only_model_v_observation_logscale}.
If it is the case, V1723 Sco could be a superbright nova.

\citet{kat26hs} calculated nova outburst cycles
for a 1.25 $M_\sun$ WD with five $\dot{M}_{\rm acc}= 5\times 10^{-11}$,
$1\times 10^{-10}$, $5\times 10^{-10}$, $1\times 10^{-9}$,
and $5\times 10^{-9} ~M_\sun$ yr$^{-1}$, and also for a 1.35 $M_\sun$ WD
with three $\dot{M}_{\rm acc}= 1\times 10^{-11}$, $5\times 10^{-10}$,
and $5\times 10^{-9} ~M_\sun$ yr$^{-1}$, as tabulated in their Table 1.

Heavy element enrichment is observed in  many classical novae 
\citep[e.g.,][]{geh98tw,hac06kb}.
To mimic such a heavy element enrichment, \citet{kat26hs} increased
carbon mass fraction of the hydrogen-rich envelope by 0.1 and decreased
helium mass fraction by the same amount at the beginning of thermonuclear
runaway \citep{chen19, sta20} from the original accreted matter of 
solar composition.
The method of their numerical calculation
is explained in \citet{kat22sha, kat24M1213, kat25hs, kat26hs}.

To fit our model $V$ light curves with the observation,
we use the $V$ luminosity of each model calculated with Equation
(\ref{free-free_flux_v-band}), that is, calculated based on
free-free emission of nova winds.
We examine V1723 Sco in more detail together with V1500 Cyg 1975
and V1674 Her 2021 based on our fully self-consistent nova outburst models.

\subsection{V1500 Cyg 1975}
\label{full_v1500_cyg_1975}


V1500 Cyg is an asynchronous polar of the orbital period $P_{\rm orb}=
0.1396$ day \citep[$=3.35$ hr, e.g., ][]{pat79}.  We adopt the distance of
$d=1.56$ kpc \citep{bai21rf} and the reddening of $E(B-V)=0.43$ 
\citep{hac26kv1674her3}.  Equation (\ref{distance_reddening_law}) gives
$(m-M)_V=12.3$.  Figure 
\ref{v1500_cyg_v1674_her_v1723_sco_full_model_fit_linear}(a) shows the $V$
(filled green squares) and Str\"omgren $y$ (filled magenta stars)
light curves of V1500 Cyg on a linear time (days after outburst) as well as
the optically-thick shocked shell model light curve (magenta line labeled
shock) calculated by \citet{hac26kv1674her3}.

Figure \ref{v1500_cyg_v1674_her_v1723_sco_full_model_fit_linear}(a) also
depicts our 1.25 $M_\sun$ WD models with five
$\dot{M}_{\rm acc}= 5\times 10^{-11}$ (orange line),
$1\times 10^{-10}$ (gray line),
$5\times 10^{-10}$ (red line),
$1\times 10^{-9}$ (blue line), and
$5\times 10^{-9}$ (black line) $M_\sun$ yr$^{-1}$,
which are taken from \citet{kat26hs}.
Note that all the model light curves are calculated with Equation
(\ref{free-free_flux_v-band}) and are depicted in the absolute $V$
magnitude. The observational $V$/$y$ data are all converted to the
absolute $V$/$y$ magnitudes with $(m-M)_V=12.3$.
Because the model $V$ light curves are plotted in the absolute $M_V$
magnitude, we can move them against the V1500 Cyg $V$ data
only in the horizontal direction.

Among the five model $V$ light curves, we adopt the orange line
of $\dot{M}_{\rm acc}=5\times 10^{-11} ~M_\sun$ yr$^{-1}$.
We add this free-free emission model light curve to Figure
\ref{v1723_sco_v1500cyg_v1674her_logscale}(a) by the black line.
Although the rising and decay phase are well reproduced
with the free-free emission model light curve (thick orange line),
the observed optical peak is much brighter than the model light curve.
\citet{hac26kv1674her3} calculated the $V$ light curve of an optically
thick shocked shell for V1500 Cyg, which is shown by the magenta line
labeled shock in Figure
\ref{v1500_cyg_v1674_her_v1723_sco_full_model_fit_linear}(a)
or by the blue line labeled shock in Figure 
\ref{v1723_sco_v1500cyg_v1674her_logscale}(a).
Note that the shocked shell illustrated in Figure \ref{wind_shock_config}
is optically thick between day 1.4 and 4 in V1500 Cyg \citep{hac26kv1674her3}.
Therefore, we see the brightness of the shell photosphere 
\citep[recombination front of hydrogen, ][]{hac26kv1674her3}. 

Our best-fit WD model of $M_{\rm WD}=1.25 ~M_\sun$ has a smallest
mass-accretion rate of $\dot{M}_{\rm acc}=5\times 10^{-11} ~M_\sun$ yr$^{-1}$
among the five models.  Therefore, the ignition mass of $8.5\times 10^{-6}
~M_\sun$ are largest among the other mass-accretion rate models 
\citep[see Table 1 of ][for ignition masses]{hac26kmaxejecta}.
\citet{hac26kv1674her3} reported that 
the mass of the ejected shell is as massive as $M_{\rm shell}\sim 
3\times 10^{-6} ~M_\sun$ near/at optical maximum.
As a result, the optical depth calculated by
Equation (\ref{potical_depth_shell}) is larger than $\tau_{\rm shell} >1$,
that is, optically thick \citep[see Figure 10a of ][]{hac26kv1674her3}.
The ejected shell has a large photosphere ($R_{\rm ph}\sim 300-600 ~R_\sun$)
that emits photons like a recombination front of hydrogen in a 
Type IIP supernova \citep[e.g.,][]{dub25}.

\citet{gal76} observed V1500 Cyg with $V$, $R$, $I$ bands, and 1.2, 1.6, 2.2,
3.6, 4.8, 8.5, 10.6, and $12.5~\mu$m bands during the 50 days
following the discovery, and concluded that the spectral energy distribution
is approximately that of a blackbody during the first 3 days while it is
close to $F_\nu =$~constant after the fourth day, where $F_\nu$ is the flux
at the frequency $\nu$.  These $F_\nu =$~constant
spectra are related to free-free emission.
\citet{enn77} obtained similar results, but based on the infrared
photometry from 1 to $20~\mu$m.  The nova spectrum changed from
a blackbody to a bremsstrahlung emission at day $\sim 4$-5,
that is, from that of a Rayleigh-Jeans tail ($F_\nu \propto \nu^2$)
to that of a thermal bremsstrahlung emission ($F_\nu \sim$ constant).
These results clearly showed that the spectrum of V1500 Cyg changed from
blackbody to free-free emission on day $\sim 4$-5.

The combination of the optically thick shocked shell \citep[magenta line, 
][]{hac26kv1674her3} and free-free emission model \citep[orange line, 
][]{kat25hs} reasonably reproduces
the light curve of V1500 Cyg.  The shocked shell luminosity dominates
the $V$ flux, at least, around the optical maximum between day 1.4 and 5.

Then, we obtain the maximum $V$ magnitude of $M_{V,\rm max}= m_{V,\rm max}
-(m-M)_V = 1.8 - 12.3 = -10.5$ and the 2 mag decay time of $t_2=2.4$ day
\citep{del20i}, where $m_{V,\rm max}=1.8$ is taken from
\citet{lock76m} and \citet{duerbeck77}.
We plot the position of V1500 Cyg in the MMRD diagram (Figure
\ref{max_t2_v1723_sco_selvelli2019_schaefer2018_all_saio_kato2026}),
the data of which are taken from \citet{del20i}.
V1500 Cyg is located above the blue line in Figure
\ref{max_t2_v1723_sco_selvelli2019_schaefer2018_all_saio_kato2026}(b),
so that V1500 Cyg is a superbright nova defined by \citet{del91}.
The superbrightness is caused by an optically thick shocked shell
at/near the $V$ maximum \citep{hac26kv1674her3}.

\subsection{V1674 Her 2021}
\label{full_v1674_her_2021}


The 2021 outburst of V1674 Her was discovered at 8.4 mag on UT 2021
June 12.537 (=HJD 2459378.037) by Seiji Ueda \citep{ueda21sk}, which
is 0.367 days after the outburst \citep[the day zero is
$t=0=t_{\rm OB}=$HJD 2459377.68; ][]{kat25hs}.
It has been observed with from radio, optical, UV, and X-ray, to gamma-ray
\citep{dra21, woo21, lin22, pat22, ori22, sok23, bha24, hab24, qui24}.

V1674 Her is an intermediate polar (IP) \citep{pat22}.
The orbital period is $P_{\rm orb}=3.67$ hr (0.1529 days) and
the spin period of the WD is $P_{\rm spin}=8.36$ min (501 s) 
\citep{mro21br, dra21, ori22, pat22, bha24, qui24}.
We adopt the distance of $d=8.9$ kpc, the reddening of $E(B-V)=0.5$,
and $(m-M)_V=16.3$
after \citet{kat25hs}, who obtained $(m-M)_V=16.3\pm 0.2$
by the time-stretching method against the three novae LV Vul, 
KT Eri, and V339 Del.

One of the remarkable features of V1674 Her is rich observational data in
the very early pre-discovery phase of the outburst (day 0.016-0.25), as shown
in Figures \ref{v1500_cyg_v1674_her_v1723_sco_full_model_fit_linear}(b)
and \ref{v1723_sco_v1500cyg_v1674her_logscale}(b),
where we assumed the origin of time to be $t=0=t_{\rm OB}=$HJD 2459377.68
after \citet{kat25hs} and \citet{hac25kv1674her2}.

In Figure \ref{v1500_cyg_v1674_her_v1723_sco_full_model_fit_linear}(b),
we plot the optical $V/g$ and visual data on a linear time (days after
outburst).  \citet{kat25hs} presented nova outburst models of
a 1.35 $M_\sun$ WD with three $\dot{M}_{\rm acc}= 5\times 10^{-9}$
(black line), $5\times 10^{-10}$ (orange line),
and $1\times 10^{-11}$ (gray line) $M_\sun$ yr$^{-1}$.
Among the three model light curves, the 1.35 $M_\sun$ WD model with
$\dot{M}_{\rm acc}= 1\times 10^{-11}$ $M_\sun$ yr$^{-1}$ reasonably
reproduce the observation.  In this model, a strong shock arises
0.32 days after the outburst and the shocked shell becomes optically thick
soon after the shocked shell is formed \citep{hac26kv1674her3}.  

The shock formation can be naturally
explained because the wind velocity decreases toward the maximum
expansion of the WD photosphere on day 0.32 but turns to increase
after day 0.32.  Then, the later ejected gas catches the earlier
ejected gas and makes a strong shock \citep{hac22k}.
The optically thick shocked shell absorbs the free-free emission
from the nova wind during from day 0.32 to day 1.0.
The shell becomes optically thin after day 1.1 and
free-free emission dominates again the $V$ luminosity.
The duration of the main optical emission mechanism is
depicted by the two-headed arrows at the bottom of Figure
\ref{v1723_sco_v1500cyg_v1674her_logscale}(b).

In V1674 Her, the first gamma-ray emission detection on day 0.39
substantially precedes the $V$ maximum on day 0.7
if an optically thick shocked shell forms after the free-free emission
$V$ luminosity reaches its maximum on day 0.32.

We obtain the maximum $V$ magnitude of $M_{V,\rm max}= m_{V,\rm max}
-(m-M)_V = 6.0 - 16.3 = -10.3$, where $m_{V,\rm max}\equiv V_{\rm max} =6.0$
is taken from the data of AAVSO.
We adopt the 2 mag decay time of $t_2=0.904$ day after \citet{hab24}.
We plot the position of V1674 Her in the MMRD diagram (Figure
\ref{max_t2_v1723_sco_selvelli2019_schaefer2018_all_saio_kato2026}).
V1674 Her is located above the blue line in Figure
\ref{max_t2_v1723_sco_selvelli2019_schaefer2018_all_saio_kato2026}(b),
so that V1674 Her is a superbright nova.
The superbrightness is also caused by an optically thick shocked shell
like in V1500 Cyg mentioned above.

\subsection{V1723 Sco 2024}
\label{full_v1723_sco_2024}

Figure \ref{v1500_cyg_v1674_her_v1723_sco_full_model_fit_linear}(c) shows the
$V$, visual, and unfiltered $V$ magnitudes of V1723 Sco on a linear timescale.
Here, we assume the outburst day to be $t_{\rm OB}=$JD 2360348.0.
The $V$ data (filled green squares) are taken from the archives of
AAVSO/the Variable Star Observers League of Japan (VSOLJ)
and a few $V$ data (filled magenta circles)
are from \citet{habtie24}.  The unfiltered $V$ (or visual) magnitudes
(filled gray stars) are from \citet{pearce24} and AAVSO.

We also depict our 1.25 $M_\sun$ WD outburst models with the same 
five $\dot{M}_{\rm acc}$ as in Section \ref{full_v1500_cyg_1975}.
Among the five model $V$ light curves, we adopt the thick blue line
of $\dot{M}_{\rm acc}=1\times 10^{-9} ~M_\sun$ yr$^{-1}$.
The observed $V$ light curve is reasonably reproduced by this model
light curve.  We depict this model light curve (black line) in
Figure \ref{v1723_sco_only_model_v_observation_logscale}.

Our best-fit WD model of $M_{\rm WD}=1.25 ~M_\sun$ has second largest
mass-accretion rate of $\dot{M}_{\rm acc}=1\times 10^{-9} ~M_\sun$ yr$^{-1}$
among the five models.  This mass-accretion rate is larger than the model
of V1500 Cyg (see Section \ref{full_v1500_cyg_1975}).
Because the ignition mass is smaller and its outburst evolution is slower, 
its shell mass is as small as $M_{\rm shell}\sim 3\times 10^{-7} ~M_\sun$
near/at the optical maximum, as already shown
in Figure \ref{v1723_sco_optical_depth_linear}.  As a result, 
the optical depth calculated by Equation (\ref{potical_depth_shell}) is
as small as $\tau_{\rm shell} <0.1$, that is, the shell is optically thin.
Free-free emission from the nova wind penetrates the shell
and dominates the optical spectrum.  
As a result, the $V$ luminosity can be well approximated by the free-free
emission calculated with Equation (\ref{free-free_flux_v-band}). 
  
For comparison,
we add the optically thick shocked shell model (magenta line labeled shock)
to Figure \ref{v1500_cyg_v1674_her_v1723_sco_full_model_fit_linear}(c) taken
from Figure \ref{v1500_cyg_v1674_her_v1723_sco_full_model_fit_linear}(a).
It does not seem that the two unfiltered $V$ data (filled gray stars) 
follow the shocked shell brightness, even if we shift the magenta line
labeled shock back and forth.  Thus, we may conclude that
the optically thick shocked shell light curve
calculated by \citet{hac26kv1674her3} for V1500 Cyg is not required.
We are able to reproduce the V1723 Sco light curve only with
our free-free emission light curve calculated with
Equation (\ref{free-free_flux_v-band}).

The position (filled blue square) of V1723 Sco in the MMRD diagram of Figure
\ref{max_t2_v1723_sco_selvelli2019_schaefer2018_all_saio_kato2026}(a)
is broadly consistent with (or close to) our free-free emission model
light curve of $M_{\rm WD}=1.25 ~M_\sun$ (blue line) with 
$\dot{M}_{\rm acc}=1\times 10^{-9} ~M_\sun$ yr$^{-1}$ (thick gray line) in
Figure \ref{max_t2_v1723_sco_selvelli2019_schaefer2018_all_saio_kato2026}(a).
Thus, we may conclude that the shocked shell in V1723 Sco is
optically thin and, as a result, V1723 Sco is not a superbright nova but
a normal nova.

Finally, we mention the shape of $V$ light curve of a nova.
The $V$ luminosity of free-free emission from a nova wind depends mainly
on the wind mass loss rate (Equation (\ref{free-free_flux_v-band})).
It should be noted that, assuming spherical symmetry, we naturally
calculated a smooth increase and then decrease of the wind mass loss rate,
$\dot{M}_{\rm wind}$, as shown in Figures \ref{v1723_sco_optical_depth_linear}
and \ref{v1500_cyg_v1674_her_v1723_sco_full_model_fit_linear}.
In these cases, the wind mass loss is continuous and the wind mass loss rate
is smoothly increasing and then decreasing.  These model $V$ light curves
reasonably reproduce the observed $V$ light curves of V1500 Cyg, V1674 Her,
and V1723 Sco.  Therefore, we may conclude that there are no multiple
mass ejections in these novae (V1500 Cyg, V1674 Her, V1723 Sco and so on).

Some novae show multiple peaks like V906 Car 2018 \citep[see, e.g.,
Figure 12 of ][]{hac22k}.  These novae suggest that the wind mass loss rate
varies significantly up and down near optical maximum.  We do not know
the reason why they change their wind mass loss rates violently.
\citet{hac22k} discussed possible reasons for such multiple peak novae.
The mass ejection rate could change up and down violently
near optical maximum if we include the effect of binarity and
the assumption of spherical symmetry is relaxed.
\citet{kat11drag} theoretically discussed such a case in more detail.

\section{Conclusions}
\label{sec_conclusion}

We have analyzed the multiwavelength light curves
of V1723 Sco and clarified the following things:\\
\begin{enumerate}
\item We have determined 
$\mu_V\equiv (m-M)_V=15.3\pm 0.2$ toward V1723 Sco
based on the time-stretching method.  The peak absolute $V$ magnitude
is calculated to be $M_{V, \rm max}= -8.5\pm 0.2$. 
\item Our nova outburst evolution model of a $1.25 ~M_\sun$ WD with 
$\dot{M}_{\rm acc}= 1\times 10^{-9} ~M_\sun$ yr$^{-1}$
reproduces well the observed $V$ light curve of V1723 Sco
by the free-free emission model of Equation (\ref{free-free_flux_v-band}),
from the early rising phase to the late supersoft X-ray source phase.
Also the supersoft X-ray (0.3-1.5 keV) light curve
observed by the Swift/XRT is reasonably explained by the blackbody
approximation of the WD photosphere with the same nova model. 
\item We regard that (1) the principal absorption line system appeared
on day 2.3 near the $V$ maximum ($=$ on day 2.0--2.5) and its expansion
velocity is $v_{\rm p}\sim 1200$ km s$^{-1}$ \citep{ayd24}
while (2) the diffuse enhanced
absorption line system arose a few days after the optical $V$ maximum
and its largest expansion velocity is $v_{\rm d}\sim 3000$ km s$^{-1}$
\citep{habtie24, sho24}.
Assuming that the velocity of the principal system is equal to the velocity
of the shocked shell and the velocity of the diffuse enhanced system is
the velocity of the inner wind, we obtain the temperature just behind
the shock to be $k T_{\rm sh}\sim 3.3$ keV, which is close to the
plasma temperature of hard X-ray 3.7 keV obtained by \citet{fau26}.
\item We have estimated the shock luminosity to be $L_{\rm sh}\sim
5\times 10^{37}$ erg s$^{-1}$ just after optical maximum based on our
1.25 $M_\sun$ WD model.  This corresponds to $L_{\rm sh}/L_{\rm opt}
\sim 0.1$ and $L_{\gamma}/L_{\rm sh} \lesssim 0.03$, which satisfies
the requirement of $\epsilon_{\rm nth}\epsilon_{\gamma}
= L_{\gamma}/L_{\rm sh} = (L_{\gamma}/L_{\rm opt})/(L_{\rm sh}/L_{\rm opt})
\lesssim 0.03$ \citep{met15fv}, because
$L_{\rm opt}\sim 7\times 10^{38}$ erg s$^{-1}$ and $L_{\gamma}/L_{\rm opt}
\sim 10^{-3.2}$-$10^{-2.5}$ \citep{fau26}.
\item Hard X-rays from optically thin plasma were detected
until day $\sim 150$, which is roughly consistent with the shock
duration estimated from our shock model until day $\sim 140$.
\item Together with our measurement of $t_2=7$ day, V1723 Sco is
located closely to an empirical MMRD relation
in the $\log t_2$-$M_{V,\rm max}$ diagram.  Therefore, this nova
is not a superbright nova, but a normal very fast nova.
This indicates that the shocked shell is optically thin because
free-free emission dominates the $V$ luminosity from the rising phase
(up)to the nebular phase. We confirmed that the shell is optically thin
($\tau_{\rm shell}\lesssim 0.1$)
based on our 1.25 $M_\sun$ WD model with $\dot{M}_{\rm acc}=1\times
10^{-9} ~M_\sun$ yr$^{-1}$.
\end{enumerate}

\begin{acknowledgments}
We are grateful to the anonymous referee for useful comments 
that improved the manuscript.
We also acknowledge with thanks the variable star observations (V1723 Sco)
from the AAVSO and VSOLJ International Database contributed by observers
worldwide
and used in this research.
\end{acknowledgments}

\vspace{5mm}
\facilities{Swift(XRT), AAVSO}


\appendix

\begin{figure*}
\epsscale{0.7}
\plotone{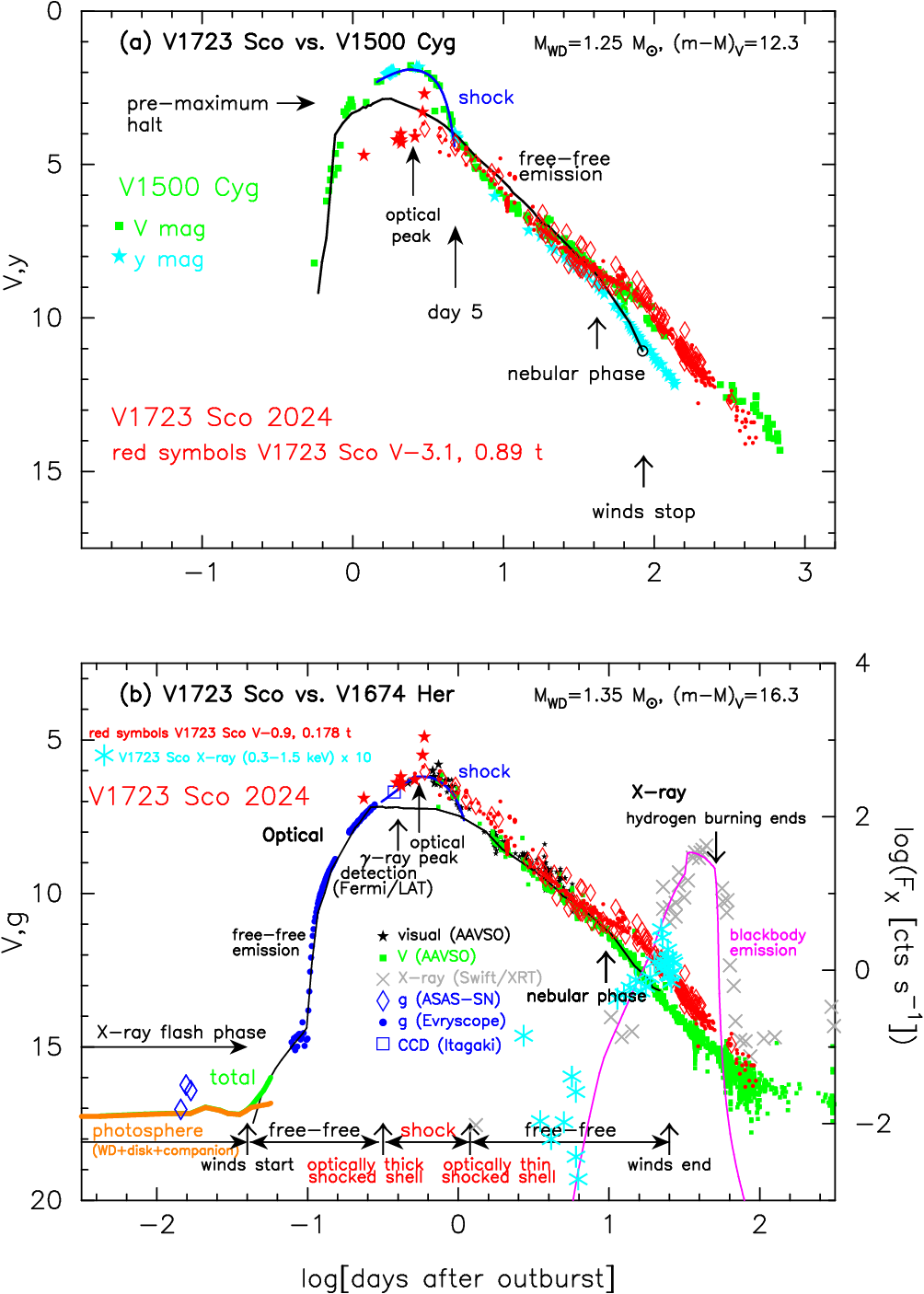}
\caption{
The $V$/unfiltered $V$ (or visual) light curve of V1723 Sco (red symbols)
against a logarithmic time.  The sources of optical data are taken from
AAVSO ($V$: filled red circles) and 
VSOLJ ($V$: open red diamonds), and \citet[][visual and unfiltered $V$:
filled red stars]{pearce24}.
The X-ray 0.3-1.5 keV data (cyan asterisks) are
from the Swift web site \citep{eva09}.
Here, we adopt the outburst day of $t_{\rm OB}=$JD 2460348.0 as day zero.
The light curve data of V1500 Cyg and V1674 Her are
all taken from Figure 7 of \citet{hac26kv1674her3}.
(a) Comparison between V1500 Cyg and V1723 Sco $V$ (or visual)
light curves.  The timescale of V1723 Sco is stretched by a factor of
0.89 ($f_{\rm s}=0.89$) and the $V$ magnitudes of V1723 Sco is
shifted up by 3.1 mag ($\Delta V= -3.1$) against those of V1500 Cyg,
as labeled by ``red symbols V1723 Sco $V$-3.1, 0.89 t''.
The black line denotes the fully-self consistent nova model of
a 1.25 $M_\sun$ WD with $\dot{M}_{\rm acc}= 5\times 10^{-11}
~M_\sun$ yr$^{-1}$, which is taken from \citet{kat26hs}.
The blue line labeled shock is an optically thick shocked shell model
for V1500 Cyg, which is taken from \citet{hac26kv1674her3}.
The black arrows and their descriptions indicate V1500 Cyg
while the red ones are for V1723 Sco.
(b) Comparison between V1674 Her and V1723 Sco $V$ (or visual)
light curves.  The timescale of V1723 Sco is stretched by a factor of
0.178 ($f_{\rm s}=0.178$) and the $V$ magnitudes of V1723 Sco is
shifted up by 0.9 mag ($\Delta V= -0.9$) against those of V1674 Her.
The blue line labeled shock is an optically thick shocked shell model
for V1674 Her, which is taken from \citet{hac26kv1674her3}.
The black line denotes our fully self-consistent nova outburst model
of a 1.35 $M_\sun$ WD with $\dot{M}_{\rm acc}= 1\times 10^{-11}
~M_\sun$ yr$^{-1}$ \citep{kat25hs}.  The main nova optical emission
mechanism of V1674 Her changes with time, which is explained
by the bottom two-headed arrows.
\label{v1723_sco_v1500cyg_v1674her_logscale} }
\end{figure*}

\section{Time-Stretching Method}
\label{appendix_time-stretching}

We have obtained $(m-M)_V=15.3\pm 0.2$ toward V1723 Sco by direct fit
of our model with the $V$ observation in Section \ref{quick_look_light_curve}.
In this Appendix, we independently determine $(m-M)_V$ by the time-stretching
method, which is a powerful way to obtain $\mu_V \equiv (m-M)_V$
toward a nova, and has ever been applied to a number of novae
\citep{hac10k, hac15k, hac16k, hac18k, hac25kv392per, hac26kv1674her3,
hac26ksbright, hac20skhs, hac24km, hac25kw, kat25hs}.

Nova light curves often show a common property; if two nova light curves
are plotted in the logarithmic time and shift in the vertical and horizontal
directions, the major part of these light curves are overlapped
each other independently of the WD mass, chemical composition, and speed
class of novae \citep{hac06kb, hac20skhs}.  Using this remarkable property,
we can determine the distance modulus to a nova
(target nova) by comparing it and a well studied nova with known
distance (template nova).

Here, we describe the $V$ light curves of the target nova as
$(m[t])_{V,\rm target}$ and the template nova $(m[t])_{V,\rm template}$.
When we adopt an appropriate time-stretching parameter $f_{\rm s}$,
these two nova $V$ light curves overlap each other.
We shift the template nova light curve in the horizontal direction
by a factor of $f_{\rm s}$ in the logarithmic scale
($t \rightarrow t\times f_{\rm s}$), and move vertically down by $\Delta V$.
This time-stretch and vertical shift can be written as
\begin{equation}
(m[t])_{V,\rm target} = \left((m[t \times f_{\rm s}])_V
+ \Delta V\right)_{\rm template}.
\label{overlap_brigheness}
\end{equation}
Then, we have the relation of
\begin{equation}
(m-M)_{V,\rm target} 
= ( (m-M)_V + \Delta V )_{\rm template} - 2.5 \log f_{\rm s},
\label{distance_modulus_formula}
\end{equation}
where $(m-M)_{V, \rm target}$ is of the target and
$(m-M)_{V, \rm template}$ is of the template \citep{hac20skhs}.
\citet{hac18k, hac18kb, hac19k, hac19kb, hac21k} confirmed that
Equations (\ref{overlap_brigheness}) and (\ref{distance_modulus_formula})
are also broadly valid for other $U$, $B$, and $I$ (or $I_{\rm C}$) bands
if each light curve is dominated by free-free emission.

This remarkable similarity is demonstrated in
Figure \ref{v1723_sco_v1500cyg_v1674her_logscale}(a), which
compares the $V$ light curve of V1723 Sco with V1500 Cyg.
The filled green squares and cyan stars denote the $V$ and 
Str\"omgren $y$ magnitudes
of V1500 Cyg while all red symbols are the $V$ or unfiltered $V$ (= clear
$V=$ CV) magnitudes for V1723 Sco.  We regard the outburst day of V1723 Sco
to be $t=0=t_{\rm OB}=$JD 2460348.0 ($=$UT 2024 February 7.5).
All the $V/y$ data of V1500 Cyg and $V/g$ (and unfiltered $V$/visual)
magnitudes of V1674 Her are all the same as those in Figures 7(a) and (b)
of \citet{hac26kv1674her3}, respectively.  The $V$ data of V1723 Sco are
taken from AAVSO (filled red circles) and VSOLJ (open red diamonds).
The visual or unfiltered $V$ (filled red stars)
are from CBET No.5346 \citep{pearce24}.

We overlap the light curves of V1500 Cyg and V1723 Sco by squeezing
the timescale of V1723 Sco by 0.89 times and shifting up the $V$ magnitude
by 3.1 mag as labeled ``V1723 Sco V-3.1, 0.89 t.''
It should be noted that we try to overlap the post-maximum phase (free-free
emission part) as long/much as possible. 
Here, we adopted the outburst day of V1500 Cyg to be
$t=0= t_{\rm OB}=$JD 2442653.0 (UT 1975 August 28.5).

In the figure, we regard V1500 Cyg as the target and V1723 Sco
as the template in Equation (\ref{overlap_brigheness}).
As V1723 Sco evolves 0.89 times faster than V1500 Cyg,
we adopt $f_{\rm s}= 0.89$ ($\log f_{\rm s} = -0.05$)
and $\Delta V= -3.1\pm0.2$ and have the relation of
Equation (\ref{distance_modulus_formula}), i.e.,
\begin{equation}
(m-M)_{V, \rm V1500~Cyg}
= (m - M)_{V, \rm V1723~Sco} -3.1\pm0.2 - 2.5 \log 0.89.
\label{distance_modulus_v1723_sco_v1500_cyg_v}
\end{equation}
Substituting $(m-M)_{V, \rm V1500~Cyg}=12.3$ \citep{hac26kv1674her3}
into Equation (\ref{distance_modulus_v1723_sco_v1500_cyg_v}),
we obtain $(m-M)_{V, \rm V1723~Sco}=15.3\pm 0.2$.

We also add our model $V$ light curve (black line) of a $1.25 ~M_\sun$ WD
with $\dot{M}_{\rm acc}= 5\times 10^{-11} ~M_\sun$ yr$^{-1}$,
which broadly reproduces the $V$ observation of V1500 Cyg
except for during the optical $V$ maximum and after the nebular phase started.
An optically thick shocked shell plays an essential role around the
optical $V$ maximum, as explained by the blue line labeled shock.
The luminosity of the shocked shell is described in more detail in
Section \ref{fully_consistent_models}.

The deviation that started after the nebular phase can be explained
by the contribution of strong emission lines such as [\ion{O}{3}].
Our model $V$ flux is calculated with free-free emission (continuum),
but does not include the effect of these strong emission lines.
The intermediate Str\"omgren $y$ band filter avoids
the region of [\ion{O}{3}] and
therefore the $y$ magnitude (filled cyan stars) can safely present
the continuum.  Thus, our model $V$ light curve broadly follows the $y$ 
light curve even in the nebular phase.

Figure \ref{v1723_sco_v1500cyg_v1674her_logscale}(b) shows V1723 Sco 
against V1674 Her.  As shown in the figure, 
V1723 Sco evolves 0.178 times faster than V1674 Her.
We adopt $f_{\rm s}= 0.178$
and $\Delta V= -0.9\pm 0.2$ and have the relation of
Equation (\ref{distance_modulus_formula}), i.e.,
\begin{equation}
(m-M)_{V, \rm V1674~Her} 
= (m - M)_{V, \rm V1723~Sco} - 0.9\pm0.2 - 2.5 \log 0.178.
\label{distance_modulus_v1723_sco_v1674_her_v}
\end{equation}
Substituting $(m-M)_{V, \rm V1674~Her}=16.3$ \citep{kat25hs}
into Equation (\ref{distance_modulus_v1723_sco_v1674_her_v}),
we obtain $(m-M)_{V, \rm V1723~Sco}=15.3\pm 0.2$.

Both equations (\ref{distance_modulus_v1723_sco_v1500_cyg_v})
and (\ref{distance_modulus_v1723_sco_v1674_her_v}) 
result in the same $(m-M)_V=15.3\pm 0.2$ toward V1723 Sco,
with the different target novae of V1500 Cyg and V1674 Her.

\end{document}